\documentclass[12pt]{article}
\usepackage{amsfonts}
\usepackage{amssymb}

\begin{document}

\title{Quantum control in infinite dimensions\thanks{%
Supported in part by Funda\c{c}\~{a}o para a Ci\^{e}ncia e Tecnologia}}
\author{Witold Karwowski\thanks{%
witoldkarwowski@go2.pl} \\
{\small Institute of Physics, University of Zielona Gora, }\\
{\small ul. Szafrana 4a, 65-246 Zielona Gora, Poland} \and R. Vilela Mendes%
\thanks{%
vilela@cii.fc.ul.pt} \\
{\small Universidade T\'{e}cnica de Lisboa and Grupo de
F\'{i}sica-Matem\'{a}tica, }\\
{\small Complexo Interdisciplinar, Av. Gama Pinto 2, 1649-003 Lisboa,
Portugal}}
\date{}
\maketitle

\begin{abstract}
Accurate control of quantum evolution is an essential requirement for
quantum state engineering, laser chemistry, quantum information and quantum
computing. Conditions of controllability for systems with a finite number of
energy levels have been extensively studied. By contrast, results for
controllability in infinite dimensions have been mostly negative,
stating that full control cannot be achieved with a finite dimensional
control Lie algebra. Here we show that by adding a discrete operation to a
Lie algebra it is possible to obtain full control in infinite dimensions
with a small number of control operators.
\end{abstract}

\section{Introduction}

To control the time evolution of quantum systems is an essential step in
many new and old applications of quantum theory\cite{Rabitz1}. Among the
fields requiring accurate control of quantum mechanical time evolution are
quantum state engineering, cooling of molecular degrees of freedom,
selective excitation, chemical reactions and quantum computing.

The first general results on controllability of quantum systems have been
obtained by Huang, Tarn and Clark (HTC)\cite{Clark}. Subsequently an
extensive amount of work has been done, mostly on establishing conditions
and degrees of controllability for systems with a finite number of energy
levels. By contrast, results for controllability in the infinite dimensional
Hilbert sphere $S_{\mathcal{H}}$ have been mostly negative, stating that
full control in $S_{\mathcal{H}}$ cannot be achieved with a finite
dimensional control Lie algebra\cite{Clark}, a result similar to the one
obtained by Ball, Marsden and Slemrod for classical control systems in
Banach spaces\cite{Marsden}.

It seemed therefore that control in infinite dimensions would require an
infinite number of distinct control operators. This is what effectively
happens in the proposal of Lloyd and Braunstein\cite{Lloyd} for quantum
computation over continuous variables. They propose, for the construction of
a universal quantum computer, the successive application of quadratic
Hamiltonians and a higher order one (for example the Kerr Hamiltonian). By
commutation, all polynomial Hamiltonians are obtained leading effectively to
an infinite dimensional control algebra.

The HTC no-go result is based on a local argument concerning the finite
dimensionality of the local manifold generated by the unitary action of the
finite dimensional control group. Left open is the question of whether the
local argument extends to a global action in the Hilbert sphere, which
however is probably true. On the other hand the result does not apply to
non-Lie groups, for example a Lie group complemented by a discrete
operation. This is the kind of situation that is explored in this paper. As
a physical motivation for the kind of control situation we deal with,
consider a charged particle in a circle (a charged plane rotator) with
Hilbert space $L^{2}(0,2\pi )$ and free Hamiltonian 
\[
H_{0}=-\frac{\partial ^{2}}{\partial \varphi ^{2}} 
\]
$\varphi \in \left[ 0,2\pi \right) $, defined in the domain 
\[
D\left( H_{0}\right) =\left\{ f\in AC^{2};f\left( 0\right) =f\left( 2\pi
\right) \right\} 
\]
where $AC^{2}$ stands for functions with absolutely continuous first
derivative. The eigenstates of $H_{0}$ are $\left\{ \left| k\right\rangle
=e^{ik\varphi };k\in \mathbb{Z}\right\} $ with eigenvalues $k^{2}$.

An application of a magnetic field pulse corresponds to the unitary operator 
\[
U_{+}=e^{i\varphi } 
\]
which shifts the eigenstates one level up 
\[
U_{+}\left| k\right\rangle =\left| k+1\right\rangle 
\]
with inverse 
\[
U_{+}^{-1}\left| k\right\rangle =\left| k-1\right\rangle 
\]

Because the energy level spacing is not uniform 
\[
\Delta E_{k}=k^{2}-\left( k-1\right) ^{2}=2k-1 
\]
one may, by resonant and non-resonant excitation, make arbitrary $U\left(
2\right) $ transformations between a particular pair of successive levels. As
it turns out (Sect. 2), these simple controls (namely $U_{+},U_{+}^{-1},U%
\left( 2\right) $) are sufficient for full controllability of this infinite
dimensional system.

\section{Control in $\ell ^{2}\left( \mathbb{Z}\right) $}

Consider the space of double-infinite square-integrable sequences 
\[
a=\left\{ \cdots ,a_{-2},a_{-1},a_{0},a_{1},a_{2},\cdots \right\} \in \ell
^{2}\left( \mathbb{Z}\right) 
\]
\[
\left| a\right| =\left( \sum_{-\infty }^{\infty }\left| a_{k}\right|
^{2}\right) ^{\frac{1}{2}}<\infty 
\]
with basis 
\[
e_{k}=\left\{ \cdots ,0,0,1_{k},0,0,\cdots \right\} 
\]
\begin{equation}
a=\sum_{-\infty }^{\infty }a_{k}e_{k}  \label{2.4}
\end{equation}

Define:

(i) A linear operator $U_{+}$ acting as a shift on the basis states 
\[
U_{+}e_{k}=e_{k+1},\qquad k\in \mathbb{Z} 
\]
and its inverse 
\[
U_{+}^{-1}e_{k}=e_{k-1},\qquad k\in \mathbb{Z} 
\]

(ii) Another linear operator $\Pi $%
\[
\begin{array}{lll}
\Pi e_{0}=e_{1} &  &  \\ 
\Pi e_{1}=e_{0} &  &  \\ 
\Pi e_{k}=e_{k} & , & k\in \mathbb{Z}\setminus \left\{ 0,1\right\}
\end{array}
\]
Then 
\[
\Pi _{n}=U_{+}^{n}\Pi U_{+}^{-n} 
\]
acts as 
\[
\begin{array}{lll}
\Pi _{n}e_{n}=e_{n+1} &  &  \\ 
\Pi e_{n+1}=e_{n} &  &  \\ 
\Pi e_{k}=e_{k} & , & k\neq n,n+1
\end{array}
\]
In the sequence $a=\sum_{-\infty }^{\infty }a_{k}e_{k}$ it exchanges $a_{n}$
with $a_{n+1}$. Likewise

\textit{Lemma 1.} Given $a\in \ell ^{2}\left( \mathbb{Z}\right) ,k\in \mathbb{Z}%
,l\in \mathbb{Z}$, the linear operator $\Pi _{k,k+l}$ $\left( l\in \mathbb{N}%
\right) $ defined by $\Pi _{k,k+1}=\Pi _{k}$ and 
\begin{equation}
\Pi _{k,k+l}a=\Pi _{k}\Pi _{k+1}\cdots \Pi _{k+l-2}\Pi _{k+l-1}\cdots \Pi
_{k+1}\Pi _{k}a  \label{2.9}
\end{equation}
for $l\geq 2$, exchanges the coefficients of $e_{k}$ and $e_{k+l}$ in (\ref
{2.4}), that is 
\[
\Pi _{k,k+l}a=a_{k+l}e_{k}+a_{k}e_{k+l}+\sum_{r\neq k,k+l}a_{r}e_{r} 
\]

\textit{Theorem 1}. Let $G\left( U_{+},\Pi \right) $ stand for the group
generated by $U_{+},U_{+}^{-1}$ and $\Pi $. Then for any $0\neq a\in \ell
^{2}\left( \mathbb{Z}\right) $ the linear span of $G\left( U_{+},\Pi \right) a$
is dense in $\ell ^{2}\left( \mathbb{Z}\right) $.

\textit{Proof}: It is sufficient to show that $b\perp G\left( U_{+},\Pi
\right) a$ implies $b=0$. Suppose $b=e_{k}$ for some $k\in \mathbb{Z}$. Since $%
a\neq 0$ there is $l\in \mathbb{N\cup }\left\{ 0\right\} $ such that at least
one of the numbers $a_{k+l}$ or $a_{k-l}$ is different from zero. Then $%
\left( b,\Pi _{k,k+l}a\right) =a_{k+l}$ or $\left( b,\Pi _{k-l,k}a\right)
=a_{k-l}$, a contradiction. Similarly if both $a$ and $b$ are terminating
sequences.

Suppose now that $b$ is terminating but $a$ is not. Let $b_{k}=0\,$for $%
\left| k\right| >N^{^{\prime }}$. Then $\exists N\leq N^{^{\prime }}$ such
that $\left( b,a\right) =\sum_{-N}^{N}b_{k}^{*}a_{k}=0$ and either $%
b_{N}^{*}a_{N}\neq 0$ or $b_{-N}^{*}a_{-N}\neq 0$. Then there is $l$ such
that $a_{N+l}\neq a_{N}$ or $a_{-N-l}\neq a_{N}$ . Hence $\left( b,\Pi
_{N,N+l}a\right) =\sum_{-N}^{N-1}b_{k}^{*}a_{k}+b_{N}^{*}a_{N+l}\neq 0$ or $%
\left( b,\Pi _{N,-N-l}a\right)
=\sum_{-N}^{N-1}b_{k}^{*}a_{k}+b_{N}^{*}a_{-N-l}\neq 0$, a contradiction.
Similarly for $a$ terminating and $b$ nonterminating.

If neither $a$ nor $b$ terminates, then there are pairs $a_{k}\neq a_{l}$
and $b_{m}\neq b_{n}$. With appropriate $g,g^{^{\prime }}\in G\left(
U_{+},\Pi \right) $ we obtain 
\[
\left( b,ga\right)
=b_{m}^{*}a_{k}+b_{n}^{*}a_{l}+b_{k}^{*}a_{m}+b_{l}^{*}a_{n}+\sum_{r\neq
k,l,m,n}b_{r}^{*}a_{r}=0 
\]
\[
\left( b,g^{^{\prime }}a\right)
=b_{n}^{*}a_{k}+b_{m}^{*}a_{l}+b_{k}^{*}a_{m}+b_{l}^{*}a_{n}+\sum_{r\neq
k,l,m,n}b_{r}^{*}a_{r}=0 
\]
Hence $b_{m}^{*}a_{k}+b_{n}^{*}a_{l}=b_{n}^{*}a_{k}+b_{m}^{*}a_{l}$, which
is possible only if either $b_{m}=b_{n}$ or $a_{k}=a_{l}$, a contradiction.

$\hspace{12cm}\blacksquare $

Now instead of the $\Pi $ operator we consider a $U\left( 2\right) $ group
operating in the linear space spanned by $e_{0}$ and $e_{1}$ and as the
identity on $\ell ^{2}\left( \mathbb{Z}\right) \ominus \left\{
e_{0},e_{1}\right\} $. In particular $\Pi \in U\left( 2\right) $.

\textit{Theorem 2}. For any $0\neq a\in \ell ^{2}\left( \mathbb{Z}\right) $ the
set $G\left( U_{+},U\left( 2\right) \right) a$ is dense in $\ell ^{2}\left( 
\mathbb{Z}\right) $.

\textit{Lemma 2}. Suppose $0\neq a\in \ell ^{2}\left( \mathbb{Z}\right) $ is a
terminating normalized sequence. Then, there is $g\in G\left( U_{+},U\left(
2\right) \right) $ such that $ge_{0}=a$.

\textit{Proof}: Let 
\[
a=a_{-N}e_{-N}+\cdots +a_{o}e_{0}+\cdots +a_{N}e_{N} 
\]
By $U\left( 2\right) $ transformations in the $\left\{ e_{0},e_{1}\right\} $
subspace and use of the $\Pi _{k,k+l}$ operators (Eq.(\ref{2.9})) one
constructs with operators $g_{i}\in G\left( U_{+},U\left( 2\right) \right) $
the following sequence 
\[
\begin{array}{ccc}
g_{1}e_{0}= & x_{1}e_{0}+a_{-N}e_{-N} & =\alpha _{1} \\ 
g_{2}\alpha _{1}= & x_{2}e_{0}+a_{-N+1}e_{-N+1}+a_{-N}e_{-N} & =\alpha _{2}
\\ 
\cdots & \cdots & \cdots \\ 
\cdots & \cdots & \cdots \\ 
g_{2N}\alpha _{2N-1}= & x_{2N}e_{0}+\sum_{-N}^{N}a_{k}e_{k} & =\alpha _{2N}
\\ 
g_{2N+1}\alpha _{2N}= & a & 
\end{array}
\]
Finally 
\[
g_{2N+1}g_{2N}\cdots g_{2}g_{1}e_{0}=a 
\]

\textit{Proof of theorem 2}: Consider $a,b\in \ell ^{2}\left( \mathbb{Z}\right) 
$ with $\left| a\right| =\left| b\right| =1$. Choose $\varepsilon $ and $N$
such that 
\[
\alpha =\left| \sum_{-N}^{N}a_{k}e_{k}\right| >1-\varepsilon 
\]
\[
\beta =\left| \sum_{-N}^{N}b_{k}e_{k}\right| >1-\varepsilon 
\]
By the lemma 2 there are $g,g^{^{\prime }}\in G\left( U_{+},U\left( 2\right)
\right) $ such that 
\[
g\sum_{-N}^{N}a_{k}e_{k}=\alpha e_{0} 
\]

\[
g^{^{\prime }}\left( \alpha e_{0}\right) =\frac{\alpha }{\beta }%
\sum_{-N}^{N}b_{k}e_{k} 
\]
Hence 
\[
\left| b-g^{^{\prime }}ga\right| \leq 2\varepsilon +\left| 1-\frac{\alpha }{%
\beta }\right| \leq 3\varepsilon 
\]

$\hspace{12cm}\blacksquare $

In conclusion: given any initial state $0\neq a\in \ell ^{2}\left( \mathbb{Z}%
\right) $ it is possible by the unitary action of an element in $G\left(
U_{+},U\left( 2\right) \right) $ to approach as closest as desired any other
state $b$ in $\ell ^{2}\left( \mathbb{Z}\right) $.

Any infinite-dimensional separable Hilbert space is isomorphic to $\ell ^{2}$%
. Therefore the results have a large degree of generality. However,
depending on the Hilbert space realization for each concrete infinite
dimensional quantum system, the control operators discussed here may or may
not be easy to implement.

\end{document}